%% file: main.tex
\PassOptionsToPackage{table}{xcolor}
\documentclass[10pt,sigconf]{acmart}
\usepackage[utf8]{inputenc}
\usepackage[T1]{fontenc}
\usepackage{lmodern}

\usepackage[table]{xcolor}
\usepackage{graphicx}
\usepackage{multirow}
\usepackage{xspace}
\usepackage{booktabs}
\usepackage{caption}
\usepackage{subcaption}

\usepackage{circuitikz}
\usepackage{tikz}
\usetikzlibrary{positioning}

\newcommand{\gcell}{\cellcolor{green!15}}
\newcommand{\bcell}{\cellcolor{blue!15}}

\newif\ifsubmission
\submissiontrue

\newif\ifanonymous
\anonymousfalse

\newif\iffullversion
\fullversiontrue

\ifsubmission
\newcommand{\mchnote}[1]{}
\newcommand{\mcnote}[1]{}
\newcommand{\arnote}[1]{}
\else
\newcommand{\mchnote}[1]{\textit{\textcolor{red}{[mch]: #1}}}
\newcommand{\mcnote}[1]{\textit{\textcolor{blue}{[marco]: #1}}}
\newcommand{\arnote}[1]{\textit{\textcolor{green}{[arnaud]: #1}}}
\fi

\newcommand{\smartparagraph}[1]{\vspace{.05in}\noindent{\bf #1}}
\newcommand{\primitivename}{\textsc{Distinct-Match}\xspace}
\newcommand{\systemname}{\textsc{Prelude}\xspace}

\copyrightyear{2018}
\acmYear{2018}
\setcopyright{acmcopyright}
\acmConference[APNet '18]{2nd Asia-Pacific Workshop on Networking}{August 2--3, 2018}{Beijing, China}
\acmPrice{15.00}
\acmDOI{10.1145/3232565.3232570}
\acmISBN{978-1-4503-6395-2/18/08}

\begin{document}
\title{\systemname: Ensuring Inter-Domain Loop-Freedom in~SDN-Enabled Networks}

\ifanonymous

\author{Paper 17}

\else

\author{\mbox{Arnaud Dethise}}
\affiliation{%
	\institution{KAUST}
}

\author{\mbox{Marco Chiesa}}
\affiliation{%
	\institution{\mbox{KTH}}
}

\author{Marco Canini}
\affiliation{%
	\institution{KAUST}
}

\newcommand{\authorreference}{Arnaud Dethise, Marco Chiesa, and Marco Canini}
\renewcommand{\shortauthors}{A. Dethise, M. Chiesa, M. Canini}

\fi

\begin{abstract}

	Software-Defined eXchanges (SDXes) promise to improve the inter-domain routing ecosystem 
	through SDN deployment. 
	Yet, the 
	na\"ive deployment of SDN on the Internet raises concerns about the 
	correctness of the inter-domain data-plane. By allowing operators to 
	deflect traffic from default BGP routes, SDN policies can 
	create permanent forwarding loops that are not visible to the control-plane.

    
    We propose \systemname, a system for detecting 
    SDN-induced forwarding loops between SDXes with high accuracy without 
    leaking private routing information of network operators.
    To achieve this, we leverage Secure Multi-Party Computation (SMPC) 
    techniques to build a novel and general privacy-preserving 
    primitive that 
    detects whether any subset of SDN rules might affect the same portion of 
    traffic without learning anything about those rules. We then leverage this 
    primitive as the main building block of a distributed system tailored 
    to 
    detect forwarding loops among any set of SDXes.
    We leverage the particular nature of SDXes to further improve the efficiency 
    of our SMPC solution.
    
    The number of valid SDN rules
    rejected by our 
    solution is 100x lower than 
    previous privacy-preserving solutions, and provides better privacy 
    guarantees. Furthermore, our solution naturally provides network operators 
    with some insights on the cost of the deflected paths.


\end{abstract}

%



\maketitle

\pagestyle{empty}

\input{body}

\subsection*{Acknowledgements}
This research is (in part) supported by European Union’s Horizon 2020 research and innovation programme under the ENDEAVOUR project (grant agreement 644960).


\bibliographystyle{abbrv}
\bibliography{main}

\iffullversion
\appendix
\section{Proof of Theorem~\ref{theo:two-sdxes-correctness-and-completeness}}

\begin{proof}
	We prove the theorem in both directions. 
	\begin{itemize}
		\item ($\Rightarrow$) We first prove that if \systemname detects a forwarding loop between two ASes at two different SDXes, then a forwarding loop exists. Suppose, by contradiction, that a forwarding loop exists between two ASes at two different SDXes but \systemname does not detect it. Let $\pi$ be the IP prefix destination towards which the deflection is created. Let $DF_1$ and $DF_2$ be the two deflections  installed at SDX$_1$ by AS$_1$ and SDX$_2$ by AS$_2$, respectively. In order for a forwarding loop to exist between AS$_1$ and AS$_2$, AS$_1$ must deflect its traffic through a BGP route that traverses AS$_2$ and vice versa. Moreover, the deflected rules must at least partially overlap in order to create a forwarding loop between AS$_1$ and AS$_1$. Yet, if the two deflected rules overlap, the \systemname detects this overlap and forbid the installation of the rule, a contradiction. 
		
		\item ($\Leftarrow$) We now prove that if there exists a forwarding loop originated because of two BGP deflections installed at two ASes at two different SDXes, then \systemname detects it. Suppose, by contradiction, that our system detects a forwarding loop but the forwarding loop does not exist. If \systemname detects a forwarding loop between two ASes, then it means that the two deflection installed at the two SDXes overlap for a certain header space of the traffic destined to $\pi$. This means that the traffic belonging to this overlap will be routed on a forwarding loop between the two ASes, which is a contradiction.  
	\end{itemize}

By observing that \systemname never learn anything about the policies of its members, the statement of the theorem is proved. 
\end{proof}
\fi

\end{document}

%% file: body.tex
\section{Introduction}\label{sec:introduction}
	In recent years, SDN has transformed the 
	way network operators design and manage networks by supporting highly 
	flexible and fine-grained network control~\cite{onf}.
%
%
	Yet, deploying SDN in practice is notoriously  
	difficult~\cite{fibbing-2015} as the co-existence 
	with legacy devices and routing protocols often leads to unpredictable 
	behaviors. 
	
	One of the most notable examples of this phenomenon has been 
	observed by Birkner et al.~\cite{sidr} in the emerging context of SDN-enabled 
	Internet eXchange Points, called Software-Defined eXchanges 
	(SDXes)~\cite{isdx}.
	SDXes have been deployed in 
	production~\cite{fiu} and are currently being tested by the largest 
	IXPs worldwide.
	Yet, na\"ive deployment of such architectures comes with 
	severe potential Internet-wide performance degradation as the growing number of 
	SDXes may increase the chances of creating inter-domain forwarding 
	loops.
	
	These unwanted forwarding behaviors stem from a simple fact: 
	there exists a mismatch between the routing expressiveness of SDN and 
	BGP, the de-facto standard inter-domain routing protocol.
	SDN at SDXes can be used to create 
	\emph{deflections}, i.e., forwarding decisions that are explicitly inconsistent 
	with the BGP forwarding decisions. Although these have the purpose to 
	optimize traffic delivery, na\"ively using them can ultimately prevent BGP 
	from guaranteeing loop-freedom. Given the practical impossibility of 
	replacing BGP in the short term with a clean-slate routing protocol, 
	SDXes must be armed with tools for avoiding forwarding loops and 
	guaranteeing a \emph{safe} (i.e., loop-free) forwarding behavior.
	
	\smartparagraph{Challenges.} Preserving the safety of Internet routing 
	is a fundamental yet complex 
	problem that entails \emph{efficiently} detecting forwarding loops 
	while \emph{preserving the privacy} of routing policies and any other information about the
	economic relationships among the networks that is deemed confidential~\cite{sixpack, 
	mpc-bgp} (see more details from a recent survey in~\cite{survey}).

	
	
	
	\smartparagraph{Our approach.} We devise \systemname,
	a novel approach to efficiently detecting forwarding loops
	that guarantees a safe forwarding behavior across the Internet, and both 
	higher levels of privacy and lower rejection rate for safe SDN policies 
	(deflections) than state-of-the-art solutions.
	
	\systemname is built upon 
	a 
	simple yet powerful network verification primitive, called \primitivename, 
	that allows any two 
	networks to verify whether their SDN rules overlap, i.e., the set of packets 
	matched by the conjunction of these rules is non-empty, without leaking 
	any information 
	about the SDN rules. We leverage recent advancements on Secure 
	Multi-Party Computation (SMPC) to evaluate functions in a 
	privacy-preserving way. To maximize efficiency, \systemname
	performs as little computation as possible within the SMPC machinery. 
	
	\smartparagraph{Our contributions. } The main contributions are: 
	\begin{itemize}
		\item We analyze the impact of BGP deflections on Internet 
		routing and prove that BGP deflections compliant with traditional yet 
		incomplete customer-provider relationships are loop-free. 
		\item We design \primitivename, a general privacy-preserving 
		primitive that allows any two networks to verify whether their SDN 
		policies overlap.
		\item We devise and evaluate \systemname, a control-plane that runs at the SDXes in 
		parallel with BGP and detects forwarding loops without leaking 
		private routing policies of the SDX members. We prove \systemname 
		identifies all forwarding loops involving any two networks connected at two distinct SDXes with \emph{zero} 
		false positives\footnote{All proofs are in the full version of our 
		paper~\cite{prelude-tr-short}.} and show that it reduces the number of false positives with more than two SDXes over previous solutions.
	\end{itemize}
	
	\smartparagraph{Related work.} The work closest to ours is 
	SIDR~\cite{sidr}, which first observed the problem of SDN-induced 
	forwarding loops and discussed how to detect them while preserving 
	privacy at the cost of accepting high false positive rates. Our solution takes 
	inspiration from other applications of SMPC for 
	IXP Route Server design, such as \textsc{Sixpack}~\cite{sixpack}, and the 
	Internet-wide Route Server~\cite{mpc-bgp}, ultimately leading to 
	almost zero false positives in detecting forwarding loops. 
    We do not rely on the SGX technology as it 
    constraints network operators to rely on very specific hardware 
    and its different and debated level of security, discussed in 
    more details in 
    ~\cite{sixpack}. We however consider SGX complementary to our 
    solution.
	Finally, 
	VeriFlow~\cite{veriflow} 
performs
	verification of SDN rules in single-domain networks that do not require 
	privacy.

%
%
%
%

\section{On IXPs and Deflections}\label{sec:background}
  In this section, we present a background on Internet eXchange Points (IXPs), 
  followed by a discussion on the problems that arise from leveraging the 
  expressiveness of SDN rules to deflect traffic away from BGP paths, a 
  problem that was introduced in~\cite{sidr}. We then 
  prove that there exists a practically relevant class of deflections that do 
  not cause any forwarding loops. 

%
	
\smartparagraph{Background on IXPs. }\label{sec:background:ixp}
	IXPs are the high-speed interconnection 
	network infrastructures through which a multitude of Autonomous Systems
	(ASes), called the IXP \textit{members}, physically interconnect
	to exchange traffic. Recently, IXPs experienced a sharp growth 
	in customer base and higher volumes of traffic due to 
	reduced path latencies and lower transit 
	costs, ultimately resulting in the existence of 
	over 800 IXPs around the world. 
	At the largest IXPs, hundreds of members exchange ever-increasing 
	volumes of traffic (in the order of $5$ Tbps) towards hundreds of 
	thousands of IP prefix destinations. 

    Routing information at IXPs is exchanged through BGP in the form of path routes 
    towards IP destination prefixes.
    Network operators specify how to rank, export, and select the routes to be 
    used for forwarding traffic through \emph{BGP policies}.
    To prevent persistent \emph{forwarding loops} (i.e., packets that are sent in 
    a network cycle), BGP uses a loop-avoidance 
    mechanism. For the scope of this paper, it is worth knowing that this 
    mechanism works as long as \emph{(i)} a single route is used to forward 
    traffic and \emph{(ii)} only this route is exported to other networks.\footnote{Any  
    stability aspect of the global Internet BGP policies is 
    orthogonal to the scope of this paper. Also, past studies have shown that 
    popular BGP destinations are remarkably stable~\cite{jrex-stability}.} 
    
	
	IXPs are deemed as a hot spot for 
	innovation~\cite{isdx, steroids}. Some of the largest 
	ones are today experimenting new SDX architectures. SDXes aim at 
	supporting advanced use cases for both IXP members and IXP 
	operators~\cite{steroids,decix-sdn} without requiring any 
	member to deploy SDN switches. Members steer their incoming\slash 
	outgoing traffic at the 
	SDX by configuring their \emph{SDN policies}, which match flows of 
	traffic (based on Layer 2 to Layer 4 fields) and apply specific actions (such 
	as forwarding a packet to a certain member)~\cite{isdx}.

	
	
	
\smartparagraph{BGP deflections at SDXes. 
}\label{sec:background:deflection}
   When IXP members install SDN policies at an SDX, the BGP 
   loop-avoidance assumptions may be violated, resulting in forwarding 
   loops~\cite{sidr}. In 
   fact, to support advanced peering applications that need to steer traffic 
   through different members according to particular QoS 
   requirements~\cite{isdx}, SDN policies support routing 
   expressiveness well beyond the IP prefix 
   destination level. Flows with the same IP prefix destination may be matched 
   by distinct SDN policies, each with a different forwarding action (e.g., one 
   policy for HTTP traffic and the other one for SSH traffic). This results in 
   some (or all) traffic being \emph{deflected} from its intended BGP path.

%
	
	\begin{figure}
		\centering
		\includegraphics[width=0.7\linewidth]{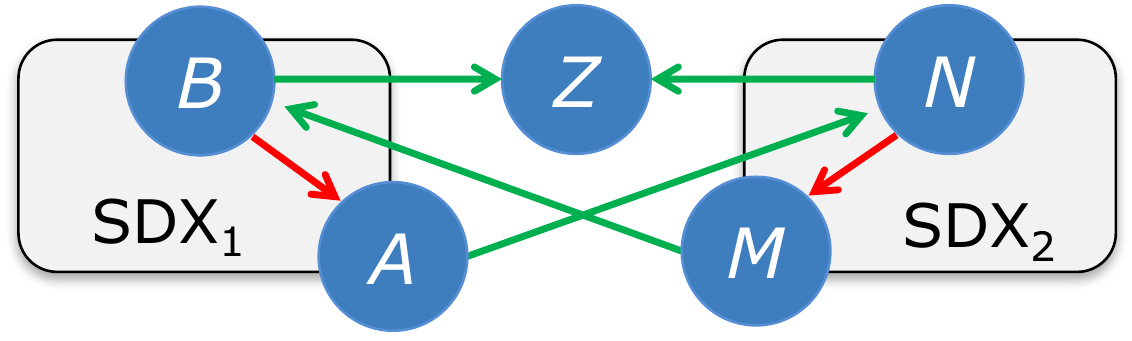}
		\caption{Example inter-SDX network. }
		\label{fig:background:loop}
	\end{figure}

	An example of an SDN-induced forwarding loop is shown in 
	Fig.~\ref{fig:background:loop}. Two networks $A$ and $B$ ($N$ and $M$) 
	connect at SDX$_1$ (SDX$_2$). A network $Z$ originates an IP prefix to 
	which all networks aim at sending traffic. Green links represent BGP paths. 
	BGP policies are such that 
	network $A$ steadily forwards traffic through path $(A\ N\ Z)$ while $M$ 
	uses $(M\ B\ Z)$. Network $B$ and $N$ install two SDN policies (depicted 
	in red) to deflect 
	their HTTP traffic (i.e., TCP port $80$) towards neighbors $A$ and $M$, 
	respectively (the order in which the rules are installed is irrelevant to the creation of a forwarding loop). This operation has a severe consequence: all HTTP traffic is 
	now forwarded through $B$, $A$, $N$, $M$, and back to $B$, a 
	forwarding loop!
	
	Detecting deflection-induced loops is difficult as deflections are not 
	propagated through BGP. Neither $B$ nor $N$ is aware of the BGP 
	deflections installed by the other and both believe the deflected path is 
	loop-free.
		
%
%
	The problem of detecting deflection-induced forwarding loops is further 
	hindered by the fact that, to quote~\cite{sidr},  \emph{``[networks] are not 
	keen 
	on sharing detailed information about their policies''}. A recent 
	survey~\cite{survey} confirmed these concerns. 
	For this reason, solutions to detect forwarding loops without disclosing 
	routing policies have been highly sought after in the recent 
	past~\cite{sixpack,mpc-bgp,sidr}. 
	The most notable one, i.e., SIDR~\cite{sidr},  
	advocates for a system where SDX members share with other SDXes  
	information about the 
	possible sets of forwarding actions of all the SDN policies affecting the 
	same specific IP prefix without disclosing the type of 
	traffic (e.g., whether it is HTTP or SSH) that is being deflected or, in other 
	words, without	sharing the SDN match filters. This solution preserves the 
	privacy of the match part of the SDN policies at the price of reporting 
	non-existing forwarding loops (false positives). For instance, if $B$'s and $N$'s deflections 
	only affect HTTP and SSH traffic, respectively, no forwarding loop is 
	created. In Sect.~\ref{sec:evaluation:lode}, we show that over one out of four 
	forwarding loops reported by SIDR are incorrect. Our system aims to both 
	accurately detect forwarding loops and improve the privacy of the 
	SDN policies.
	
    \smartparagraph{Safe deflections exist! }One may ask if it is possible to install BGP deflections in such a way that Internet routing is 
    loop-free without requiring any coordination among ASes. It is well-known 
    that when BGP policies are configured according to traditional 
    customer-provider economic relationships, as defined by the well-known 
    Gao-Rexford conditions~\cite{gaorexford}, Internet 
    routing converges to a loop-free forwarding state. We generalize this to include BGP deflections created by SDN policies. 
	
	\begin{theorem}\label{theo:gao-rexford-deflections}
		If the BGP policies and deflections satisfy the Gao-Rexford conditions, 
		then the forwarding state is loop-free.
	\end{theorem}
	
	Unfortunately, while Gao-Rexford conditions are often implemented in 
	practice, there are still many exceptions to them such as partial transit 
	relationships~\cite{bgp_survey}. Moreover, customer-provider
	relationships are generally confidential. Thus, we need ways to detect
	potential forwarding loops before deflections are installed.

	

	

\section{\systemname}\label{sec:design}
	We are now ready to discuss the system requirements that an Internet 
	loop-detection system should meet to be of practical interest.
	We will then explain how we carefully design \systemname to tackle the challenges
	of satisfying such requirements and carry out loop detection in a distributed
	manner (Sect~\ref{sec:design:overview}), and present the privacy-preserving
	primitive \primitivename for detecting overlaps of SDN rules
	(Sect~\ref{sec:design:primitive}). 
	

	\smartparagraph{System requirements. }
	Based on the discussion from Sect.~\ref{sec:background}, we argue that 
	any mechanism for detecting 
	deflection-induced forwarding loops among SDXes should satisfy the 
	following four requirements: \emph{(i)} \emph{safety}, i.e., no packet is forwarded 
	through a cycle, \emph{(ii)}
	\emph{privacy}, i.e., SDN policies should be kept private to 
	the largest possible extent and only information that is strictly necessary to ensuring the safety of routing or available through BGP should be revealed, \emph{(iii)} \emph{effectiveness}, i.e., reject as few safe 
	SDN policies as possible (ideally zero), and \emph{(iv)} \emph{performance}, i.e., 
	forwarding loops 
	should be detected as soon as possible. We make extensive use of recent 
	advances in computation over encrypted data (i.e., SMPC)
	to achieve these four properties. 
	
		\smartparagraph{Threat model. } 
		We rely on the \textit{curious-but-honest} security model, in 
		which an attacker aims to 
		learn as much information as possible 
		about an SDX member's policies from the exchanged messages while 
		adhering to the agreed protocol. This assumption resonates well in 
		today's IXP 
		ecosystem~\cite{sixpack} as SDX members already trust their IXPs\slash 
		SDXes for forwarding their traffic to the intended destinations. 
		An \textit{attacker} is either an SDX operator or an SDX member who 
		wants to learn about the SDX 
		policies of a member of another SDX. 
		We assume that the attacker does not gain control of any two SDXes, 
		which are independent organizations.
        SDXes may learn about the IP prefix destinations of SDN policies 
        installed at remote SDXes that deflect traffic through them but nothing 
        more (neither who installed them nor the type of traffic involved). 
        These privacy assumptions are stronger than the ones defined in
        SIDR~\cite{sidr} as the actions of the SDN policies are hidden from third parties.
	
	\smartparagraph{Root-causes of forwarding loops. }We observe that 
	only two types of operations can trigger a 
	forwarding loop: \emph{(i)} installation\slash removal of an SDN policy or \emph{(ii)} BGP 
	updates to the routing state (e.g. a path change\slash withdrawal). 

\subsection{Overview}\label{sec:design:overview}
\begin{figure}
		\centering
		\includegraphics[width=0.7\linewidth]{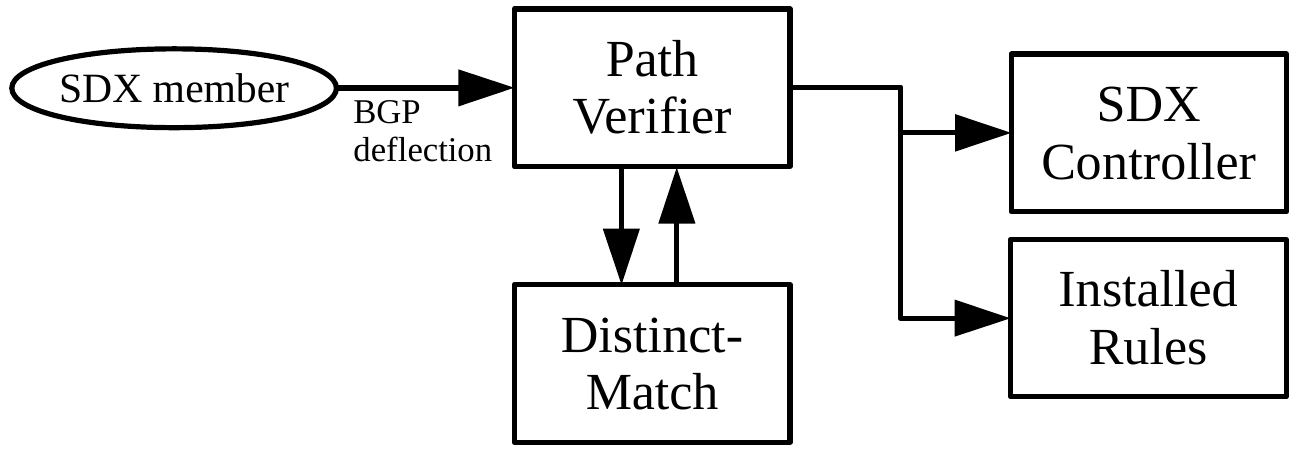}
		\caption{\systemname overview. When a request is received, \systemname verifies the safety of the
		deflected path. If activation is allowed, the rule is saved in the local
		table.}
		\label{fig:design:overview}
	\end{figure}

	Fig.~\ref{fig:design:overview} shows the high-level architecture of \systemname.
	
	\smartparagraph{The path verifier. }
	When a member of an SDX installs\slash removes an SDN policy or there is 
	a BGP update that impacts a deflection, the member sends a 
	request to the 
	SDX where the rule is being installed. 
%
	This request is received by the 
	\emph{path-verifier} running at the SDX 
	connected to the issuing member.
	The path-verifier is 
	responsible for determining which SDXes are traversed by a given path and 
	how a path would be deflected. 

	
	To function properly, the path-verifier relies on a primitive called \primitivename. This primitive will indicate whether two distinct SDN rules (or policies) affect the same traffic.
	Through \primitivename, the SDX can verify, for a given flow of packets and associated path, if any other SDX along the path is deflecting the traffic.
	If a deflection occurs, the system will verify whether we are closing a loop. \primitivename is implemented in a privacy-preserving way, and the SDX can verify the traffic path without learning \textit{anything} about the SDN policies of the requesting member.


	
	All SDXes contacted during the verification will record the installation of a rule and notify the installing SDX of future downstream changes in policies or BGP state.

	\smartparagraph{Interactions with the SDX controller. }
	\systemname interacts with an SDX \emph{controller}, which is responsible
	for installing SDN policies. \systemname filters unsafe policies
	before they get installed into the SDX forwarding tables.
	
	\smartparagraph{Provable effectiveness guarantees. }
	A common case is to have just two SDXes announcing the same prefix and potentially exchanging traffic with each other.
	In this case, our system guarantees provable effectiveness, as formalized in Theorem~\ref{theo:two-sdxes-correctness-and-completeness}. 	If more than two SDXes are involved, false positives can occur (more in Sect. \ref{sec:evaluation}).
			
	\begin{theorem}\label{theo:two-sdxes-correctness-and-completeness}
		\systemname detects a forwarding loop between two ASes at two SDXes if and 
		only if a forwarding loop between the two ASes exists. Moreover, the members' SDN policies 
		are not revealed neither among SDXes or any unintended member. 
	\end{theorem}
    
    \iffullversion
\begin{proof}
The proof of the theorem is straightforward. Similarly to~\cite{gaorexford}, we model the network as a graph $G=(V,E)$ where each vertex in $V$ represents an AS and each edge in $E$ represents a BGP session among two ASes. We say that a directed edge $(u,v)$ is a \textit{customer-to-provider} (\textit{peer-to-peer}) edge if $u$ is a customer (peer) of $v$. Similarly, we say that a directed edge $(u,v)$ is a \textit{provider-to-customer} edge if $u$ is a provider of $v$.
First, observe that BGP routing is guaranteed to reach a stable state when the Gao-Rexford conditions hold~\cite{gaorexford}. This allows us to focus only on the existence of forwarding loops at the data-plane level. Observe that, when Gao-Rexford conditions hold, each path is guaranteed to be \textit{valley-free}, i.e., each path consists of a sequence of customer-to-provider edges, followed by possible one peer-to-peer edge, possibly followed by a sequence of provider-to-customer edges. Now, observe that any Gao-Rexford compliant deflection cannot alter these sequence, i.e., a deflection can modify the forwarding path but it cannot alter the customer-to-provider, peer-to-peer, and provider-to-customer composition. Hence, a forwarding loop cannot arise, which proves the statement of the theorem. 
\end{proof}

\fi
	
	\smartparagraph{Example. } In the example topology from figure 
	\ref{fig:background:loop}, suppose that AS~$B$ has already installed a 
	policy $r_B$ at SDX$_1$. AS~$N$ requests the installation of policy $r_N$ 
	at SDX$_2$, using BGP path $(M\ B\ Z)$. Upon reception of this request, 
	the path-verifier at SDX$_2$ discovers the presence of SDX$_1$ on the 
	path. \primitivename is called to discover if $r_N$ conflicts 
	with any policy at SDX$_1$  (i.e. the set of packets matched by the policies 
	is non-empty). If that is the case, SDX$_1$ notifies SDX$_2$ that the next 
	deflection on the deflected path is AS~$N$ at SDX$_2$. The loop between 
	SDX$_1$ and SDX$_2$ is detected, and the rule is rejected.

\subsection{The \primitivename primitive}\label{sec:design:primitive}
		
	We assumed that the \primitivename primitive used by the path-verifier to detect if two rules will affect the same traffic can be implemented \emph{without} revealing to that SDX the rule identifying the traffic flow.
	We now provide some essential background on SMPC, then describe the specific rule representations and SMPC ``circuits'' we used to \textit{efficiently} implement \primitivename.
	
	\smartparagraph{Secure Multi-Party Computation. }
	SMPC is a cryptographic technique that allows multiple parties to jointly compute the outcome of a function \textit{f} while keeping their respective inputs (and outputs) private. 
	
	
	In our work, we rely on two well-established SMPC protocol implemented in the ABY~\cite{aby} framework. The first one is Goldreich-Micali-Wigderson (GMW), in which the core idea is to evaluate the successive gates sequentially. The second is the Yao's garbled circuits protocol, which relies on garbled truth tables for the circuit gates. Both protocols have different performance behavior that are discussed in Sect. \ref{sec:evaluation}.
	
	In both protocols, the function \textit{f} is represented as a boolean circuit consisting of basic AND and NOT gates that is jointly evaluated by \emph{two} non-colluding entities. The inputs to the circuit are XOR-encrypted before being shared by the parties, thus ensuring perfect privacy.
	In our model, each of the non-colluding SMPC entities executing the protocol is an SDX.
	
	Efficiency requires that the SMPC function be kept as simple as possible to limit the number of rounds in which entities need to exchange information. To maintain practical runtimes, we limited our use of SMPC to \textit{overlap detection}, as described next.
	
	
	\smartparagraph{\primitivename SMPC execution.} Consider SDX$_1$,
	that wants to verify whether a rule $r_A$ overlaps 
	with a remote rule $r_B$ installed at a remote SDX$_2$. 
	Each SDN policy (or rule) is encoded as a string in which each bit has three 
	possible 
	values: $1$, $0$ or $x$ (don't care). Two rules are said to be \textit{distinct} 
	or \textit{non-overlapping} if there exists a bit index such that one rule 
	matches a $1$ and the other rule matches a $0$. In SMPC, we model each 
	rule $r_i$ as a bit pattern $p$ and a bit mask $m$.
	In the mask, the $i$-th bit 
	$m_i$ is 1 if $r_i$ is different from $x$, $0$ otherwise. In the pattern, the 
	$i$-th bit $p_i$ has the same value as $r_i$, unless $r_i$ is $x$, in which case 
	it can be any value.
	
	The SMPC execution consists of three phases: input generation, circuit 
	execution, and output reconstruction.
	Each SMPC input consists of two shares obtained by 
	XOR-encoding the input with a random nonce: one share is the random 
	nonce and the other is the encoded string. Without both shares, it is 
	impossible to learn anything about the actual input.  Each SDX sends to 
	the other SDX only one share of the input.
		
	\begin{figure}
		\centering
		\begin{subfigure}[b]{0.35\linewidth}
			\centering
			\resizebox{\textwidth}{!}{\begin{circuitikz} \draw
				(0, -0) node[and port, rotate=270] (bothcare) {}
				(2, -0) node[xor port, rotate=270] (differnt) {}
				(1, -2) node[and port, rotate=270] (distinct) {}
				(bothcare.in 2) node[anchor=south] (m1) {$m_1$}
				(bothcare.in 1) node[anchor=south] (m2) {$m_2$}
				(differnt.in 2) node[anchor=south] (p1) {$p_1$}
				(differnt.in 1) node[anchor=south] (p2) {$p_2$}
				(bothcare.out) -| (distinct.in 2)
				(differnt.out) -| (distinct.in 1)
				(distinct.out) node[anchor=north] (res) {$c =$ 1 if disjoint bit, 0 otherwise}
				;
			\end{circuitikz}}
			\caption{Masked XOR}
			\label{fig:design:overlap-mxor}
		\end{subfigure}%
		~
		\begin{subfigure}[b]{0.35\linewidth}
			\centering
			\begin{tikzpicture}
				\node[rectangle, draw, inner sep = 7pt, text badly centered, fill = white, thick] (mux) at (0,0) {MUX};
				
				\node (vi) [above = 0.5cm of mux.130] {$v_i$};
				\node (vd) [above = 0.5cm of mux.50] {$v_d$};
				\node (di) [left = 0.5cm of mux.180] {$d_i$};
				\node (ui) [below = 0.5cm of mux.270] {$u_i$};
				
				\draw[->] (vi.south) -- (mux.130);
				\draw[->] (vd.south) -- (mux.50);
				\draw[->] (di.east) -- (mux.180);
				\draw[->] (mux.270) -- (ui.north);

				\draw (vi.south) ++(-0.08 , -0.15) -- ++(0.16, 0.1);
				\draw (vd.south) ++(-0.08 , -0.15) -- ++(0.16, 0.1);
				\draw (mux.270) ++(-0.08 , -0.15) -- ++(0.16, 0.1);
			\end{tikzpicture}
			\caption{Value mapper}
			\label{fig:design:overlap-mux}
		\end{subfigure}
		\caption{SMPC boolean circuits of \primitivename.}
	\end{figure}
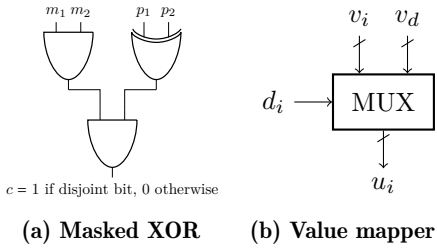
	
	Figure \ref{fig:design:overlap-mxor} shows the boolean circuit implementing the SMPC function that compares the rules $r_1$ and $r_2$. For each bit, this circuit returns $1$ if the bits of the corresponding index in each rule match $0,1$ or $1,0$, respectively (i.e., if the rules are disjoint). We then connect all the bits with an OR gate to return a single (encrypted) bit for the comparison.
	
	If the rules are non-distinct (i.e., some packets would be matched by both 
	rules), it means that the remote SDX already has an active overlapping rule 
	that would deflect some traffic of the new SDN rule.
	In that case, the circuit returns the identifier of the next SDX in the deflection path created by the rule installed at SDX$_2$.
	If the rules are distinct, a special value is returned as illustrated in Figure \ref{fig:design:overlap-mux} ($v_i$ is the next-hop identifier for rule $i$, and $v_d$ stands for ``dummy'').
	The SDXes finally exchange their shares and reconstruct the output.
	
	To make the verification efficient, \primitivename compares 
	multiple (potentially thousands of) rules in parallel and returns a set of all the 
	next-hops created by the active rules. The output of the circuit is randomly 
	and secretly shuffled to protect the remote SDX's rules.
	
	Section \ref{sec:evaluation} explains the cost of using this primitive when more than two SDXes are involved in a loop.
	
	\smartparagraph{Implementation.} We implemented and evaluated the 
	above boolean circuits using the ABY framework~\cite{aby}, a C++ toolbox 
	supporting various implementations of SMPC. The circuits can be 
	evaluated with either the GMW or Yao's garbled circuit protocol. Both 
	protocols return the same result, but have performance differences that are 
	discussed in Sect.~\ref{sec:evaluation:distinct-match}.
	
	Because the bottleneck for SMPC circuit evaluation is the one-way 
	communication delay (and not the local computation), the queries to 
	\primitivename are done in parallel for each SDX --- only successive 
	deflections need to be discovered sequentially. This means that the cost 
	to verify the safety of a new rule is equal to the cost of the longest path 
	in the deflection graph.

\section{Evaluation}\label{sec:evaluation}
	We now assess the effectiveness of \systemname in 
	terms of its ability to correctly identify safe or unsafe SDN policies. Prior to that, we show the results of our 
	micro-benchmarks for the \primitivename primitive in terms of processing 
	time for different number of rules, which is the costliest part of our system. 
	We finally expand our evaluation to the performance of the whole system. 
	We also investigate how different types of BGP deflections may affect 
	the length of the routing paths.


	\begin{table}
		\centering
		\small
		\begin{tabular}{ll|c|c|c|c|c}
			\hline
			\multicolumn{2}{c|}{} & \multicolumn{4}{c|}{Number of rules} & Baseline\\\hline
			\multicolumn{2}{c|}{Delay} & 1 & 50 & 500 & 5000 & 5000 \\
			
			\hline
			\multirow{2}{*}{1 ms}
			& setup	 & \bcell 2.75 & \bcell 12.6 & \gcell 94.5 & \gcell 1033 & —   \\
			& online & \bcell 4.63 & \bcell 10.2 & \gcell 30.5 & \gcell 151 & 2   \\
			
			\hline
			\multirow{2}{*}{10 ms}
			& setup	 & \bcell 20.7 & \bcell 48.4 & \bcell 162 & \gcell 1086 & —   \\
			& online & \bcell 40.6 & \bcell 64.2 & \bcell 136 & \gcell 382 & 20  \\
			
			\hline
			\multirow{2}{*}{100 ms}
			& setup	 & \bcell 201  & \bcell 408  & \bcell 1237  & \gcell 6061 & —   \\
			& online & \bcell 401  & \bcell 604  & \bcell 1395  & \gcell 2638 & 200 \\
			
			\hline
		\end{tabular}
		\caption{Average computation time (in ms) for the comparisons. Blue (Green) cells use Yao (GMW).}
		\label{table:evaluation:distinct-match}
	\end{table}

\smartparagraph{Distinct-Match micro-benchmarks. }\label{sec:evaluation:distinct-match}
	We evaluate the performance of \primitivename. 
	In this micro-benchmark, two SMPC entities jointly evaluate the 
	\primitivename circuit. One SMPC entity holds \emph{one single} SDN rule 
	and the other holds a \emph{set} of SDN rules. 
	We study the performance of our \primitivename implementations along 
	three different dimensions: \emph{(i)} the time to evaluate the SMPC circuit, \emph{(ii)} the 
	communication delay between the two entities, and \emph{(iii)} the size of the set of 
	SDN rules. We break down evaluation time by \textit{setup} 
	time, which is a phase of SMPC computation that can be pre-computed and depends only on the maximal number of possible rules, and \textit{online} 
	time, which depends on the actual inputs and directly affects the system 
	reactiveness. 

	
	Table \ref{table:evaluation:distinct-match} shows the average execution 
	time over 50 executions of the \primitivename operation for different 
	communication delays and sizes of the set of rules. In all experiments, we 
	set the length of the rules to 13 bytes, which is sufficient to encode 
	source\slash destination IPv4\slash TCP addresses\slash ports.
	The baseline is $2 * RTT$, ignoring any local computation cost. 
    Evaluating the circuit with Yao's garbled circuit protocol (blue cells) is more 
    efficient for small numbers of rules but scales linearly in the number of rules.
    Using the GMW protocol (green cells) scales 
    logarithmically with the number of rules and, depending on the delays, is more efficient when we 
    compare hundreds or thousands of rules at once.
    We also observe that the online time is directly proportional to the 
    round-trip delay, clearly indicating that the communication delay among the 
    two SMPC entities is the limiting factor that prevents achieving 
    even more performing SMPC executions.

\smartparagraph{Loop-detection effectiveness. }\label{sec:evaluation:lode}
	We now compare the ability of \systemname to effectively installs as many 
	SDN policies as possible with SIDR~\cite{sidr}, the state-of-the-art 
	mechanism. To evaluate it, we used the same simulation technique 
	from SIDR. The topology is built from the CAIDA 
	AS graph~\cite{caida} augmented with the links from the combined IXP 
	dataset. It computes paths to 1000 randomly selected 
	AS destinations.
	Each of the 421 IXPs is considered to be an 
	SDX. Moreover, we assume that traffic traverses an SDX if its 
	path traverses two members of that IXP one after 
	the other. 
	
	Each SDX member generates between one and four policies towards 20\% of the other members (at most 50 members). Those policies are restricted to matching the IP protocol (TCP or UDP) and a random source or destination port.
	Unlike SIDR, we select the source and destination port randomly. This represents the fact that no rule will be present for the most commonly used ports, as the policies for those are implemented through BGP. This also avoids creating duplicate rules. As a consequence, the number of created loop is significantly lower and those loops are often more complex than those considered in the SIDR evaluation.
	A policy from member $A$ to member $B$ is applied to each route announced by $B$.
	
\begin{figure}
		\centering
		\includegraphics[width=0.8\linewidth]{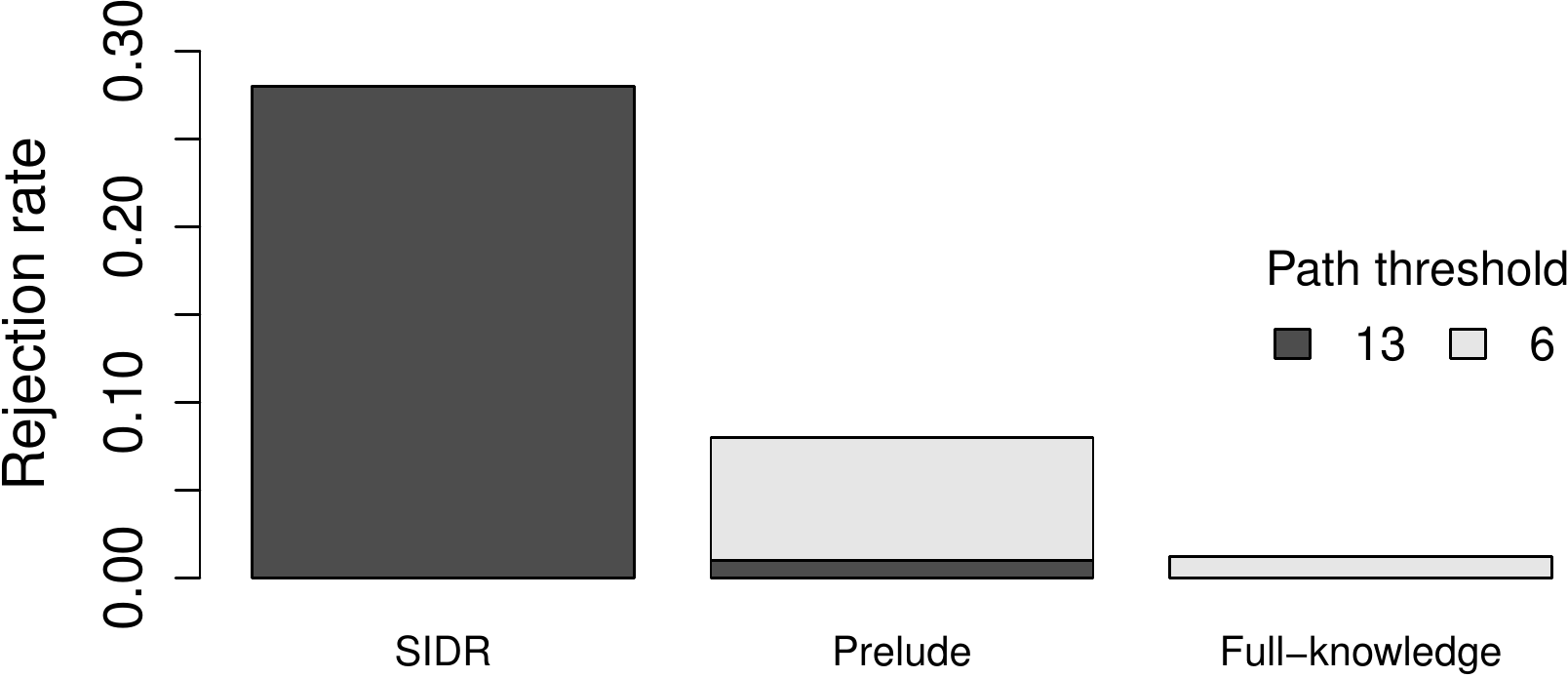}
		\caption{Fraction of correct rules rejected by various approaches. 
		}

		\label{fig:evaluation:fprate}
	\end{figure}

	Figure \ref{fig:evaluation:fprate} shows the results for three loop 
	detection systems: SIDR, \systemname,
	and a perfect knowledge approach based on path exploration without any privacy 
	restrictions, which we use as an ideal upper bound for our approach. We 
	look at 
	a path exploration dimension called 
	\emph{path-threshold}, which caps the maximum number of deflections to 
	be followed by a path exploration mechanism.  We observe that the false 
	positive rate of SIDR in our setting 
	(random ports) is at 28\% while the false positive rate of the perfect 
	knowledge solution is 0\% (with a path-threshold of 13).
	
    We observe that \systemname sharply reduces the number of incorrectly detected 
    loops when compared to the less privacy-preserving SIDR. With a 
    path-threshold of $6$ SDXes, \systemname has only $8\%$ of 
    false positives, which drops to $0.3\%$ when the path-threshold is set to 
    $13$.

\begin{figure}
		\centering
		\includegraphics[width=0.7\linewidth]{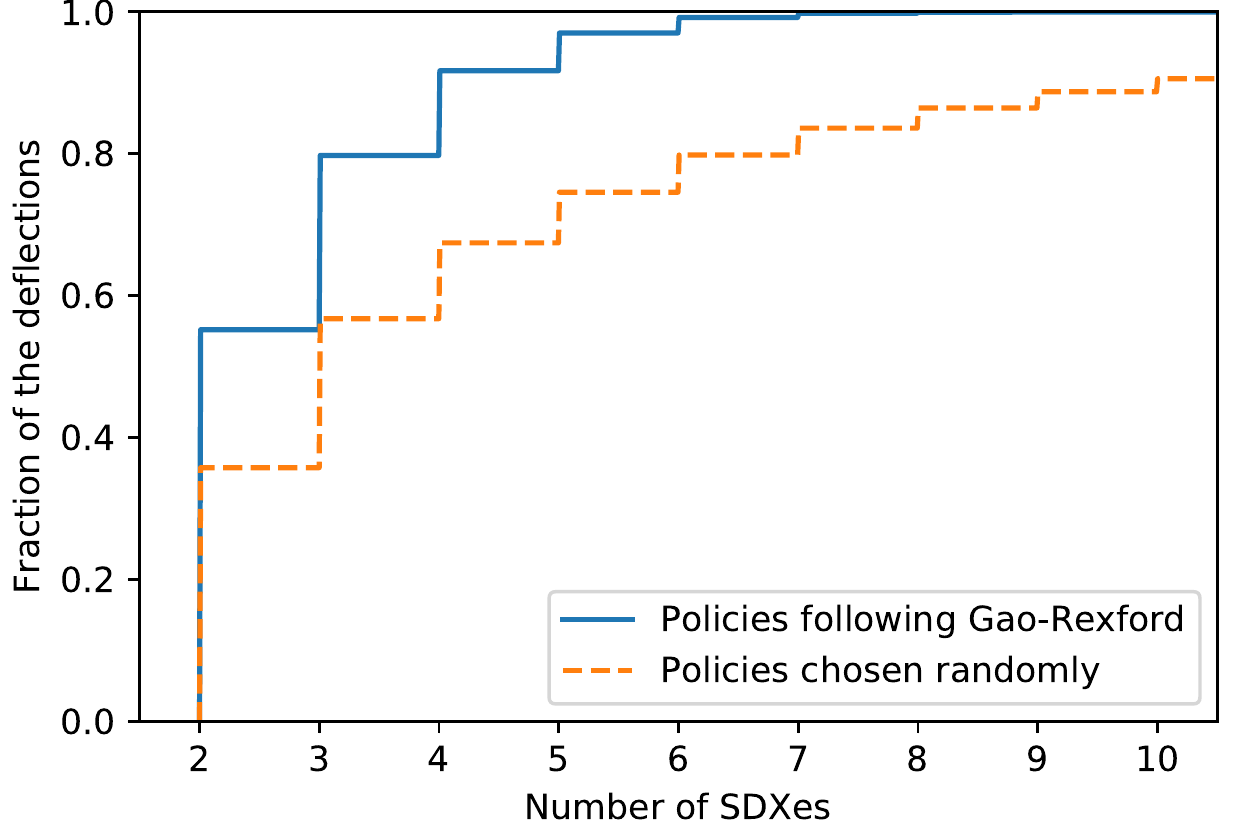}
		\caption{Number of SDXes in the deflected paths.
		}
		\label{fig:design:pathlen}
	\end{figure}

\smartparagraph{Path exploration performance. }
	The time required for verification is proportional to the number of SDXes 
	traversed by the deflected path and the \primitivename execution time. 
While exploring the network hop-by-hop is normally costly, real paths 
generally cross only a few ASes~\cite{bgp_data}, and similarly only cross a 
very limited number of SDXes. We evaluated the number of SDXes crossed 
by deflected paths and report the CDF in Figure~\ref{fig:design:pathlen}. We consider two classes of BGP deflections:  
Gao-Rexford (GR) compliant ones~\cite{gaorexford} (blue solid line), i.e., an 
AS never deflects traffic to a provider if a customer route is available and non-compliant ones
(orange dashed line), chosen randomly. 
We observe that GR conditions keeps the number of SDXes in the 
path low (60$\%$ of the paths with only two SDXes). Yet, GR conditions are 
deemed non realistic in 
practice as exceptions may exist. 
The second 
scenario represents a worst-case situation, where ASes deflect traffic through 
any random route they learned, resulting in higher path inflation. We observe 
that the SIDR assumption about the presence of an SDX in a path is an 
over-estimate on the number of traversed SDXes.
When only two SDXes are traversed by a deflected path, the 
verification cost is exactly equal to the one of executing \primitivename.

%
	
\section{Conclusion}
	In this paper, we aimed at reconciling SDXes with the limited BGP routing expressiveness and ASes' natural desire for preserving the privacy of their business information.
	We designed and implemented a general privacy-preserving primitive for detecting overlaps among SDN rules that can be executed in practical runtimes.
	We then introduced \systemname, a framework that allows SDXes to verify the safety of SDN policies while keeping those policies private.
	Finally, we investigated the impact of SDN deflections on Internet routing, demonstrating positive conditions for loop-free Internet routing

%